# Durability of MgO/hydromagnesite mortars – Resistance to chlorides and corrosion

Fabio Enrico Furcas[1*], Alexander German[1,2], Frank Winnefeld[2], Pietro Lura[1,2], Ueli Angst[1]

[1] ETH Zürich, Institute for Building Materials, 8093 Zürich, Switzerland

[2] Empa, Swiss Federal Laboratories for Materials Science and Technology, Laboratory for Concrete and Asphalt, 8600 Dübendorf, Switzerland

[*]corresponding author (Email: ffurcas@ethz.ch)

## Abstract

The durability of MgO/hydromagnesite mortars was studied with respect to their corrosion performance and resistance to chloride attack and moisture. MgO/hydromagnesite pastes were cured in chloride solution to induce potential formation of Mg-chlorides; however, no such phases were observed. Rapid chloride ingress measurements demonstrated high penetration resistance and low chloride migration coefficients, i.e. $D_{Cl} = 1 \times 10^{-13}$ to $1 \times 10^{-12}$ m$^2$/s. The corrosion rate of carbon steel embedded in MgO/HY mortars, as determined by linear polarization resistance measurements, was in the range $i_{corr} = 1 \times 10^{-9}$ A/cm$^2$ in dry and $1 \times 10^{-7}$ A/cm$^2$ in wet conditions, irrespective of the mortar composition or curing condition. These findings corroborate the hypothesis that, in the absence of chlorides, the moisture condition is the primary predictor of corrosion rate of carbon steel in the MgO/hydromagnesite binder. These accelerated, short-term experiments suggest that the binder may be suited to protect embedded carbon steel from corrosion under specific exposure conditions of practical relevance.

# 1    Introduction

MgO/hydromagnesite binders have been recently investigated as an alternative to traditional Portland cement clinker with potentially low environmental footprint [1-4]. Their main component, reactive MgO, can be produced from Mg-silicates [5-7] or Mg-rich brines [8, 9], thereby eliminating the intrinsic $CO_2$ emissions associated with the calcination of $MgCO_3$. Hydration of MgO in presence of hydromagnesite ($Mg_5(CO_3)_4(OH)_2 \cdot 4H_2O$ – abbreviated as HY) leads to formation of a *hydrous carbonate-containing brucite* (abbreviated as HCB), which, being the sole hydration product in the MgO/HY binder, is responsible for its mechanical properties [3, 10]. The sustainable sourcing of MgO notwithstanding, the environmental footprint of cementitious products made from MgO/HY further depends on their long-term durability. In particular, in the most important case of structural concrete reinforced with steel, the corrosion performance of carbon steel embedded in concrete produced from these novel binders is paramount.

Chloride-induced corrosion is recognized as one of the major causes for the structural degradation of reinforced concrete structures [11]. In addition to the local breakdown of steel passivity [12, 13], chloride ingress into reinforced MgO/HY concrete may further induce phase changes including the formation of soluble or expansive Mg-chlorides such as bischofite ($MgCl_2 \cdot 6H_2O$) or phases typical for Mg-oxychloride cements (Sorel phases)[1] [14]. Whilst Mg-oxychloride cements are generally deemed incompatible with steel reinforcements due to their low pore solution pH and high chloride concentrations [15], the formation of these phases in MgO/HY concrete may be positive, in that it may limit the free chloride concentration at disposal to catalyze steel corrosion. However, to the best of our knowledge, no results about the durability of MgO/HY binder exposed to chloride-rich environments have been reported thus far in the literature.

In comparison to traditional Portland cements, the pore solution of MgO/HY binder has a low pH of approx. 10.5-11.0 [3]. From a durability viewpoint, this comparatively moderate degree of pore solution alkalinity is considered a pitfall, as the protective film of iron (hydr)oxide phases on the surface of the steel reinforcement ceases to be thermodynamically stable [16]. In contrast to the prevailing doctrine that the loss in reinforcement passivity inevitably leads to accelerated corrosion [17], a number of publications [18-21] as well as practical case studies [22] show that the moisture content is a significantly stronger predictor of the reinforcement corrosion rate. The question as to whether the pore solution pH of MgO/HY binder is, analogous to the corrosion of steel in carbonated Portland cement concrete, tolerable, as long as the moisture condition can be controlled, must be further investigated.

---

[1] Hydration products of Mg-oxychloride cements have variable stoichiometry depending on mix compositions and curing temperature: $x$ $Mg(OH)_2 \cdot y$ $MgCl_2 \cdot z$ $H_2O$, e.g. $x$-$y$-$z$ = 3-1-8 (3' phase), 5-1-8 (5' phase), etc.

This paper investigates the durability of the MgO/HY binder with regard to its resistance against chloride attack, the ingress of moisture and corrosion performance. The resistance against chloride attack was investigated by (i) phase analyses of pastes cured in chloride solution and (ii) rapid chloride ingress tests of mortars. Linear polarization resistance (LPR) and single-frequency impedance measurements provide further insights into the corrosion performance of the MgO/HY binder under the capillary uptake of (i) water and (ii) concentrated chloride-containing solution. Two different MgO/HY mortar mixes with MgO-to-HY ratios of 90/10 and 70/30 by mass were prepared for the experiments. To study the effect of curing atmosphere on the corrosion rate, samples prepared for electrochemical measurements were further cured at two relative humidities (RH), i.e. 57% and 98%. Findings are discussed with particular emphasis on the compatibility of the MgO/HY binder with steel reinforcement.

## 2   Materials and methods

### 2.1   Materials

The MgO/HY binder was mixed from commercially available raw materials. The reactive MgO powder was a calcined product of natural magnesite of medium reactivity with a specific surface area (SSA) of 31.1 $m^2$/g measured by nitrogen gas adsorption (range of SSA of medium reactive MgO: 10-60 $m^2$/g [23]). The HY was a natural mix of HY and huntite ($CaMg_3(CO_3)_4$). The mineral composition of the natural mix was determined by Rietveld refinement, yielding 67% HY, 25% huntite, 7% dolomite ($CaMg(CO_3)_2$), and 1% other minor phases, e.g. quartz. Both the X-ray diffraction patterns and the chemical composition of the reactive MgO and the huntite-dolomite mix are appended to this study as electronic supplementary materials (compare Figures A1 and A2, Table A1). The particle sizes were measured with a laser particle size analyzer (Malvern Mastersizer X, Germany), using isopropanol for dispersion (compare Figure A3). The binder raw materials were homogenized via a Turbula mixer in a 1 L plastic bottle for 2 h.

MgO/HY pastes for phase analysis were prepared with a w/c of 0.50 and the addition of 3.5% polycarboxylate ether (PCE) superplasticizer (solid content: 32 %). Pastes were mixed with a vacuum mixer and filled into cylindrical plastic vessels (inner diameter 33 mm), sealed and cured at 20°C for 24 h. Afterwards, samples were taken out of the plastic vessels and cut into 5.0-5.5 mm-thick slices with an approximate mass of 7-8 g. Mortars for corrosion experiments and rapid chloride ingress tests were prepared with a sand-to-binder ratio of 3:1 by mass, using an aggregate mixture (Normensand GmbH,

Beckum, Germany) conforming to DIN EN 196-1. The mortars were mixed according to the mixing procedure described in EN 196-1. After mixing, mortars were filled into molds and compacted by vibration. Mortars were cured at 20°C and 98% RH for 24 h, demolded and transferred into their dedicated curing environments, as described in the following method sections. Prior to mixing both the pastes and mortar samples, the PCE superplasticizer was dissolved in deionized water.

## 2.2 Methods

## 2.1 Phase analyses of pastes

Phase analyses of MgO/HY pastes was performed on 90/10 and 70/30 paste slices. After cutting, paste slices were pre-conditioned at 20°C and 98% RH for 27 d (Stage I). Subsequently, they were cured in separate plastic vessels filled with 0.2 M KOH + 3% NaCl solution at 20°C for 7 and 28 d, i.e. same solution as used for rapid chloride ingress tests (Stage II). Reference pastes were analyzed directly after curing at 20°C and 98% RH for 28 d, without being exposed to alkaline or chloride solution. To discriminate the effect of pH and chlorides, a separate set of pastes was cured in an alkaline solution (0.2 M KOH) without any NaCl addition to solely see the effect of pH on the formation of hydration phases. After curing, paste slices were crushed and hydration was stopped by organic solvent exchange [24]. The pastes were then ground below 0.063 mm for subsequent XRD and TGA measurements. Table 1 summarizes paste compositions and curing conditions of pastes prepared for phase analysis.

*Table 1:* MgO/HY pastes prepared for phase analysis. All samples were prepared with a w/c ratio of 0.50 and 3.5% SP concentration.

| MgO-to-HY mixing ratio | Stage I, pre-conditioning | | Stage II, curing | |
|---|---|---|---|---|
| | Relative humidity | Curing time | Solution | Curing time |
| 90/10 (reference) | 98% | 27 d | - | 28 d |
| 90/10 | 98% | 27 d | 0.2 M KOH | 7 d, 28 d |
| 90/10 | 98% | 27 d | 0.2 M KOH + 3% NaCl | 7 d, 28 d |
| 70/30 (reference) | 98% | 27 d | - | 28 d |
| 70/30 | 98% | 27 d | 0.2 M KOH | 7 d, 28 d |
| 70/30 | 98% | 27 d | 0.2 M KOH + 3% NaCl | 7 d, 28 d |

X-ray diffraction patterns were measured in an X'Pert Pro (Malvern Panalytical, UK) in Bragg-Brentano geometry. The diffractometer uses a scanning line detector X'Celerator (Malvern Panalytical, UK), Cu K$\alpha_1$ radiation ($\lambda$ = 1.54059 Å) and a Ge (111) Johansson monochromator. All samples were scanned across a range of 2$\theta$ = 5 – 75° in steps of 0.0167° 2$\theta$.

Ground paste powders were analyzed using an STA 449 F3 Jupiter TGA instrument (Netzsch, Germany). Approximately 50 to 60 mg of the powders were filled into an alumina crucible ($Al_2O_3$, mass

= 325 mg) and heated under a nitrogen atmosphere from 30 °C to 1000 °C at a heating rate of 10 K/min. The nitrogen gas flow rate was 20 ml/min.

## 2.2 Rapid chloride ingress test on mortar samples

The rapid chloride ingress test was performed according to the Swiss standard SIA 262/1 Appendix B [25], which specifies that a Portland cement-based concrete of certain exposure classes must pass an accelerated chloride ingress test. Resistance in regard to chloride attack is assessed by the chloride migration coefficient $D_{Cl}$, which must not exceed $10 \cdot 10^{-12}$ m$^2$/s for exposure classes XD2b(CH) and XD3(CH) to pass the test[2] [25].

90/10 and 70/30 MgO/HY mortars for rapid chloride ingress tests were filled in 150×150×150 mm$^3$ cube molds. The cube molds were demolded after 24 h of curing at 98% RH and then cured for another 19 d. After a total curing time of 20 d, three cylindrical samples of 50 mm diameter and length were drilled from the cubes. Subsequently, the cylindrical samples were cured for another 7 d in deionized water and then prepared for testing (after 28 d in total). Before the measurement, the cores were cleaned in an ultrasonic bath for 120 s and dried with a cloth. Then, a latex foil was sealed around the cylindrical samples with the original surface and its opposite surface remaining uncovered by the foil. The samples were placed into separate measurement devices, which were cured in alkaline chloride solution (0.2 M KOH + 3% NaCl). The measurement device consisted of a metal cage, with an anode and a cathode linked to a power source (DC) and a cylinder filled with a chloride-free, alkaline solution (0.2 M KOH) (compare Figure A4). The two opposing, uncovered sides of the cylindrical samples were brought into contact with the chloride-free, alkaline solution (side of the anode) and the alkaline chloride solution (side of the cathode). The measurements were started by applying 20 V to induce chloride migration towards the anode. Current, voltage, and temperature of both solutions were monitored and if necessarily adjusted during the measurements. After 24 h the measurement was stopped and the samples were taken out, dried and split in halves to examine the longitudinal cross-section. To reveal chloride migration fronts, cross-sections of samples were dyed with 0.1% fluorescein solution (ethanol as solvent) and 0.1 N silver nitrate solution. Areas into which the chlorides had penetrated were colored pink by the dyeing procedure. The samples were then dried at 40°C for another 24 h. After drying, the pink color in chloride-rich areas was bleached out, while areas with no chlorides appeared much darker. Chloride penetration depth was measured at six equidistant points for each cross-section of a sample. The depth was measured from the sample's surface, which was in contact with the chloride solution to the chloride migration front. Wet bulk density $\rho_{wet}$ and chloride migration coefficient $D_{Cl}$ were determined using Eq. 1 and Eq. 2, respectively.

---

[2] For exposure class XD1(CH), the chloride ingress test after SIA 262/1 Appendix B should be substituted by the carbonation resistance test and obtained results should not exceed the limit of 5.5 mm/y.

$$\rho_{wet} = \frac{4m_w}{d^2\pi h} \qquad \text{Eq. 1}$$

with $m_w$ = sample mass after storage in water [kg], $d$ = sample diameter [m], $h$ = sample height [m].

$$D_{Cl} = \frac{z}{t} \cdot (x_d - 1.5462) \cdot \sqrt{z \cdot x_d} \ [m^2/s] \qquad \text{Eq. 2}$$

with $z = 8.619 \cdot 10^{-5} \frac{hT}{U}$ [m] ($T$ being average temperature of the alkaline chloride-free and chloride-rich solutions [K], $U$ = applied voltage [V]), $t$ = measuring time [s], $x_d$ = mean chloride ingress depth [m].

## 2.3 Electrochemical measurements on mortar samples

The corrosion performance of carbon steel in MgO/HY mortars was studied using an experimental setup shown in Figure 1. The specimen design and the placement of embedded working and counter electrodes was further optimized to allow for the simultaneous monitoring of moisture ingress through capillary suction, as detailed in Schmid et al. [26]. As schematically illustrated in Figure 1, carbon steel rods and pairs of stainless steel rods were embedded in mortar prisms (150×25×60 mm³) at a cover depth of 10, 20 and 30 mm relative to the sample base.

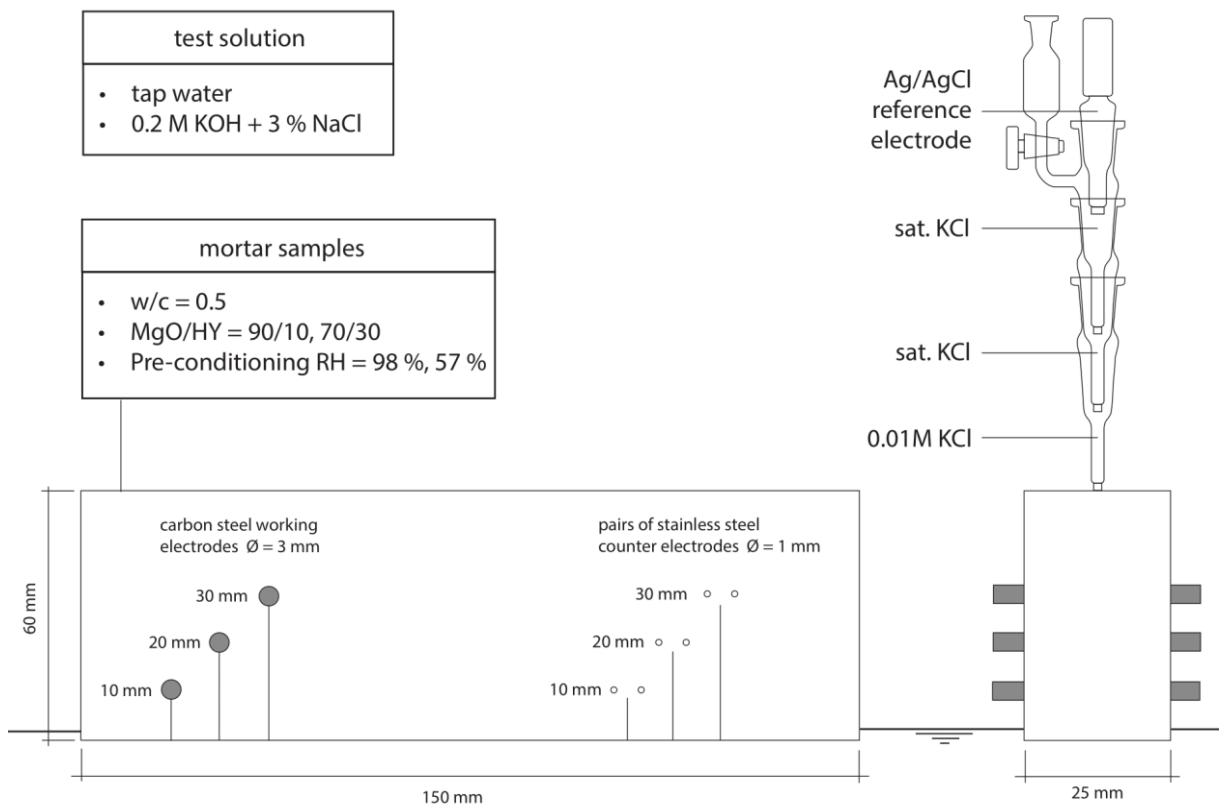

*Figure 1: Schematic illustration of the electrochemical setup to monitor the corrosion rate of carbon steel in MgO/HY mortar samples exposed to the capillary ingress of (i) tap water and (ii) 0.2 M KOH + 3 mass-% NaCl.*

After pre-conditioning at 98% or 57% RH for 27 days, the prisms were placed in a test solution reservoir of (i) tap water and (ii) 0.2 M KOH + 3 mass-% NaCl solution, and the ingress of electrolyte perpendicular to the water table was monitored in a series of single-frequency two-point impedance measurements between the pairs of equidistant stainless steel rods. One mortar prism was used for each test solution and curing condition. Impedance measurements were performed continually for 3 to 5 days in a 5-minute interval, at a frequency of 1000 Hz and an AC amplitude of 1 V, using a data logger and sensor system provided by the company DuraMon (Zürich, Switzerland). The instantaneous corrosion rate of carbon steel embedded in MgO/HY mortars was further evaluated by a series of linear polarization resistance (LPR) measurements in a 3-electrode setup. For each embedded carbon steel working electrode (WE), the stainless steel rod closest to the WE at the same cover depth was used as counter electrode (CE). Working electrodes underwent potentiodynamic polarization from -20 to +20 mV relative to the open circuit potential (OCP), and then back down again to -20 mV in cathodic direction at a scan rate of $v$ = 0.167 mV/s. The instantaneous current density, $i_{corr}$, was computed according to

$$i_{corr} = \frac{b_a \times b_c}{\ln(10) \times (b_a + b_c)} \times \frac{1}{R_p} \; [\text{A/cm}^2],$$  **Eq. 3**

where $R_p$ is the polarization resistance in $\Omega \, \text{cm}^2$ and $b_a$ and $b_c$ are the anodic and cathodic Tafel slopes in V [27], where $\frac{b_a \times b_c}{b_a + b_c} = 12$ mV.

All measurements used a silver-silver chloride (Ag/AgCl$_{sat}$) reference electrode (RE) in saturated KCl manufactured by Willi Möller AG (Zürich, Switzerland). Before each experiment, the RE potential was calibrated and the measured potentials corrected against a laboratory-internal reference-reference electrode. The RE was assembled into a stack of three porous frit glass capillaries, the two topmost capillaries containing saturated KCl and the bottom one containing diluted 0.01 M KCl, to ensure a stable reference potential and prevent the leakage of highly concentrated chloride solution into the mortar sample. The RE stack was placed on top of the mortar prisms, as illustrated in Figure 1. Corrosion and moisture ingress measurements were performed for different sample compositions (90/10 and 70/30 MgO/HY) and curing conditions (98 % and 57 % RH). As previously summarized in Table 1, all samples were cured for 28 days at 20 °C.

# 3    Results

## 3.1    Phase analyses of pastes cured in chloride solution

### 3.1.1    TGA

Figure 2 and 3 show TGA results of 90/10 and 70/30 reference pastes after pre-conditioning and pastes cured for further 7 and 28 d in alkaline chloride solution. Formation of HCB was evident in both reference pastes from the high mass loss between 300-430°C linked to loss of $H_2O$ and $CO_2$ [3]. Mass loss of HCB in the 70/30 reference was lower due to the smaller amount of formed HCB. In addition, a decomposition peak of unreacted HY was measured at around 440°C in the 70/30 reference. Minor decomposition peaks of huntite (stepwise decarbonation) were measured at around 560°C and between 630-700°C (630-730°C for the 70/30 reference). Mass loss due to gel water from HCB was observed between 30-200°C for both reference pastes [3].

Pastes, which were cured for 7 and 28 d in alkaline chloride solution, did not show any additional, separate decomposition peak related to newly formed phases, e.g. Mg-chlorides. However, such peaks, if present, could possibly be masked by larger decomposition peaks of HCB, unreacted HY or huntite. The mass loss related to decomposition of HCB between 300-430°C was higher for pastes cured in alkaline chloride solution. However, curing in alkaline chloride solution resulted in a strong shift of the main HCB decomposition peak towards lower temperatures. A comparison of TGA data of pastes cured in the alkaline chloride-free solution and in the alkaline chloride solution (Figures A5 and A6 in the ESM) revealed the shift of the decomposition peak of HCB to be the result of chloride additions to the curing solution. While reference pastes and pastes cured in alkaline solution showed no shift of the HCB peak, pastes cured in alkaline chloride solution exhibited a significant shift. Furthermore, curing in solutions with and without chlorides had an effect on the amount of gel water released during the TGA measurement between 30 and 200°C. Pastes cured in alkaline chloride solution exhibited in general lower amounts of gel water than the reference pastes.

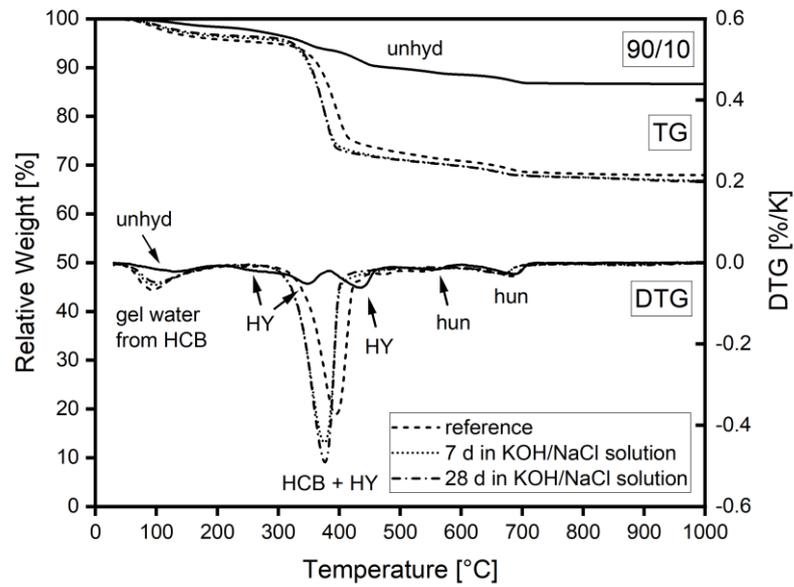

*Figure 2:* TGA data of an unhydrated 90/10 binder and a 90/10 paste cured at 98% RH for 28 d (reference before curing in alkaline chloride solution) and after curing in alkaline chloride solution (0.2 M KOH + 3% NaCl) for 7 d and 28 d. Various DTG peaks are labeled and abbreviated as hydromagnesite (HY), huntite (hun) and hydrous carbonate-containing brucite (HCB).

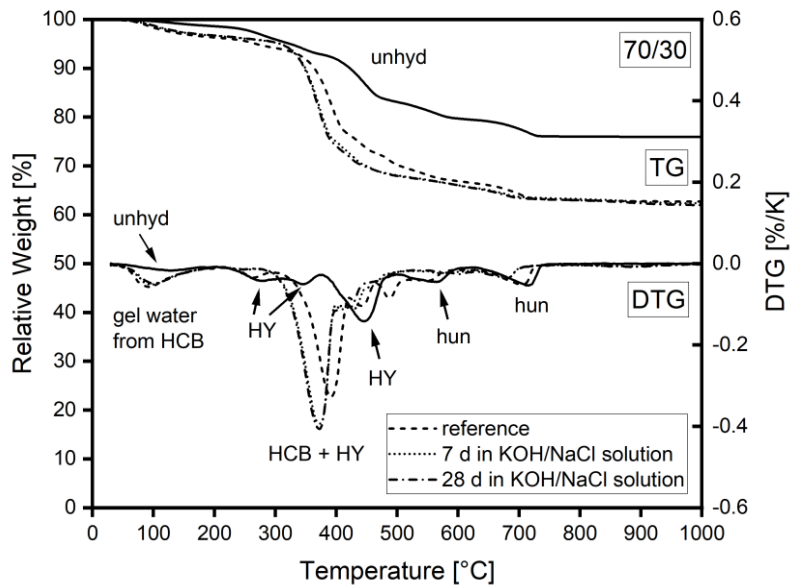

*Figure 3:* TGA data of an unhydrated 70/30 binder and a 70/30 paste cured at 98% RH for 28 d (reference before curing in alkaline chloride solution) and after curing in alkaline chloride solution (0.2 M KOH + 3% NaCl) for 7 d and 28 d. Various DTG peaks are labeled and abbreviated as hydromagnesite (HY), huntite (hun) and hydrous carbonate-containing brucite (HCB).

### 3.1.2  XRD

Figures 4 and 5 show diffraction patterns of 90/10 and 70/30 pastes before and after curing in alkaline chloride solution, respectively. Pre-curing of pastes at 98% RH and 20°C for 27 d after cutting led to hydration of MgO and HY resulting in the formation of HCB. The HCB reflections showed characteristic broadening and 001 reflection splitting as a result of stacking faults. The formation of Mg-chlorides or any other additional phases besides HCB was not observed in 90/10 and 70/30 pastes cured for 7 and 28 d in alkaline chloride solution. Thus, the presence of chlorides in solution did not affect the general phase assemblage of the MgO/HY pastes. Curing in alkaline chloride solution led only to further hydration of MgO and HY, resulting in formation of more HCB. These observations are in agreement with the results obtained from TGA, demonstrating a higher mass loss related to decomposition of HCB between 300 and 430°C, as a function of the hydration time. Figures A7 and A8 in the ESM compare 90/10 and 70/30 pastes cured in alkaline solution (0.2 M KOH) and alkaline chloride solution. The pastes did not show any significant difference with regard to phases found across all binder compositions and exposure conditions.

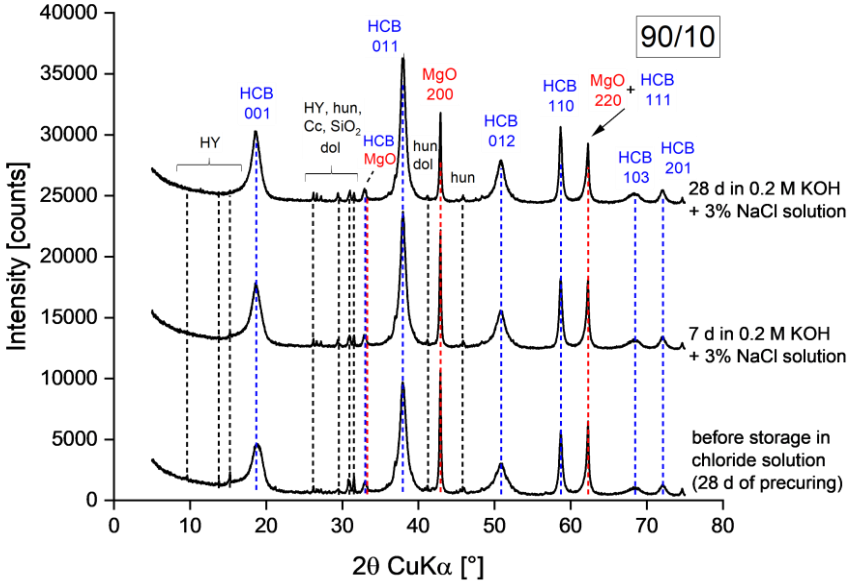

*Figure 4:* Diffraction patterns of a 90/10 paste cured at 98% RH for 28 d (reference before curing in alkaline chloride solution) and after curing in alkaline chloride solution (0.2 M KOH + 3% NaCl) for 7 d and 28 d. The main peak positions of various mineral constituents are abbreviated as hydromagnesite (HY), hydrous carbonate-containing brucite (HCB), magnesium oxide (MgO), huntite (hun), dolomite (dol), silicon dioxide ($SiO_2$) and calcium carbonate (Cc).

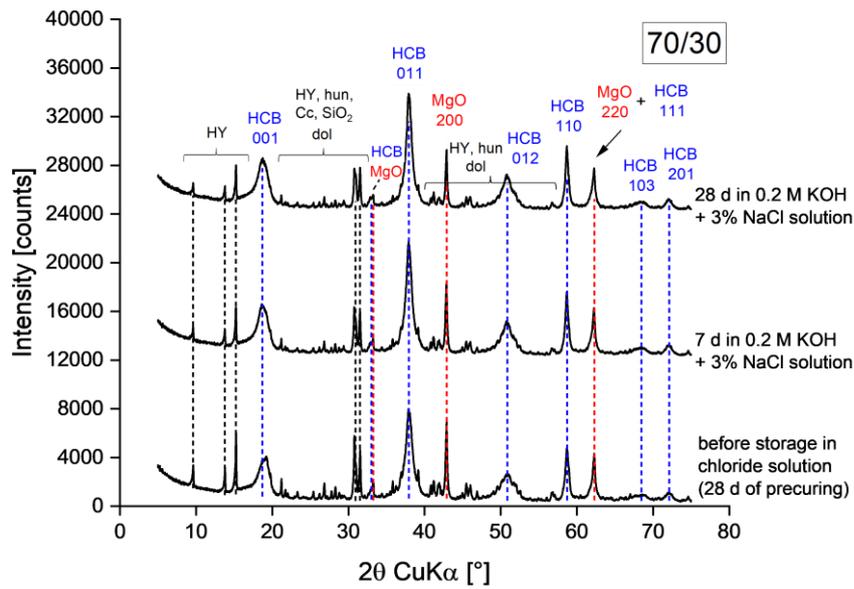

*Figure 5:* Diffraction patterns of a 70/30 paste cured at 98% RH for 28 d (reference before curing in alkaline chloride solution) and after curing in alkaline chloride solution (0.2 M KOH + 3% NaCl) for 7 d and 28 d. The main peak positions of various mineral constituents are abbreviated as hydromagnesite (HY), hydrous carbonate-containing brucite (HCB), magnesium oxide (MgO), huntite (hun), dolomite (dol), silicon dioxide ($SiO_2$) and calcium carbonate (Cc).

## 3.2 Rapid chloride migration test

Mortars of both binder compositions showed only marginal chloride penetration depths between 1-5 mm (total sample length: 50 mm), with mean penetration depths of around 2-3 mm (see **Error! Reference source not found.** and **Error! Reference source not found.** for 90/10 and 70/30 MgO/HY mortars, respectively). Graphical representations of chloride penetration profiles for 90/10 and 70/30 mortars are displayed in Figures A9 and A10 in the electronic supplementary materials (ESM), respectively. Figure A11 in the ESM shows exemplary pictures of 90/10 mortar halves, split and dyed. Mean chloride migration coefficients $D_{Cl}$ were similar for both samples: $(0.6 \pm 0.3) \times 10^{-12}$ and $(1.0 \pm 0.6) \times 10^{-12}$ m$^2$/s for the 90/10 and 70/30 mortars, respectively.

*Table 2:* Chloride ingress test results of the 90/10 mortar.

|  | sample #1 | sample #2 | sample #3 | mean |
|---|---|---|---|---|
| diameter [mm] | 49.5 | 49.6 | 49.5 | 49.5 ± 0.0 |
| wet bulk density [kg/m³] | 2199 | 2190 | 2202 | 2197 ± 7.0 |
| applied voltage [V] | 20.3 | 20.4 | 20.3 | 20.3 ± 0.1 |
| mean chloride ingress depth [mm] | 1.3 | 0.8 | 1.9 | 1.3 ± 0.6 |
| max. chloride ingress depth [mm] | 2 | 3 | 4 | 3.0 ± 1.0 |
| chloride migration coefficient $D_{Cl}$ [m²/s] | $0.6 \times 10^{-12}$ | $0.3 \times 10^{-12}$ | $1.0 \times 10^{-12}$ | $(0.6 \pm 0.3) \times 10^{-12}$ |

*Table 3:* Chloride ingress test results of the 70/30 mortar.

| | sample #1 | sample #2 | sample #3 | mean |
|---|---|---|---|---|
| diameter [mm] | 49.5 | 49.4 | 49.5 | 49.5 ± 0.1 |
| wet bulk density [kg/m³] | 1972 | 1992 | 1968 | 1977 ± 13.0 |
| applied voltage [V] | 20.0 | 20.2 | 20.2 | 20.2 ± 0.0 |
| mean chloride ingress depth [mm] | 3.1 | 1.2 | 1.7 | 2.0 ± 1.0 |
| max. chloride ingress depth [mm] | 5 | 2 | 4 | 3.7 ± 1.5 |
| chloride migration coefficient $D_{Cl}$ [m²/s] | $1.7 \times 10^{-12}$ | $0.5 \times 10^{-12}$ | $0.8 \times 10^{-12}$ | $(1.0 \pm 0.6) \times 10^{-12}$ |

## 3.3 Corrosion experiments of steel rebars embedded in MgO/HY mortars

Figures 6 to 9 show various two-point impedance measurements as a function of the sample composition, exposure time, cover depth and curing conditions. Across all data series obtained, the normalized impedance measurements reduced the fastest for the pair of stainless steel rods closest to the water table, that is, at a cover depth of 10 mm, followed by those embedded at 20 and 30 mm. Samples made from 90/10 MgO/HY responded equally rapid to the ingress of water, irrespective of the curing condition (Figures 6a and 7a). Due to the high electrolyte conductivity, the ingress of highly concentrated chloride-containing solution (Figures 6b and 7b) prompted a more significant drop in the measured impedances than the ingress of tap water (Figures 6a and 7a). At a cover depth of 30 mm, the observed reduction in the measured impedance appears insignificant across the entire time span of the experiment. Prisms cast with 70/30 MgO/HY were significantly less resistant to the capillary ingress of water, as evident from a faster and more pronounced reduction in the measured impedances compared to their 90/10 counterparts.

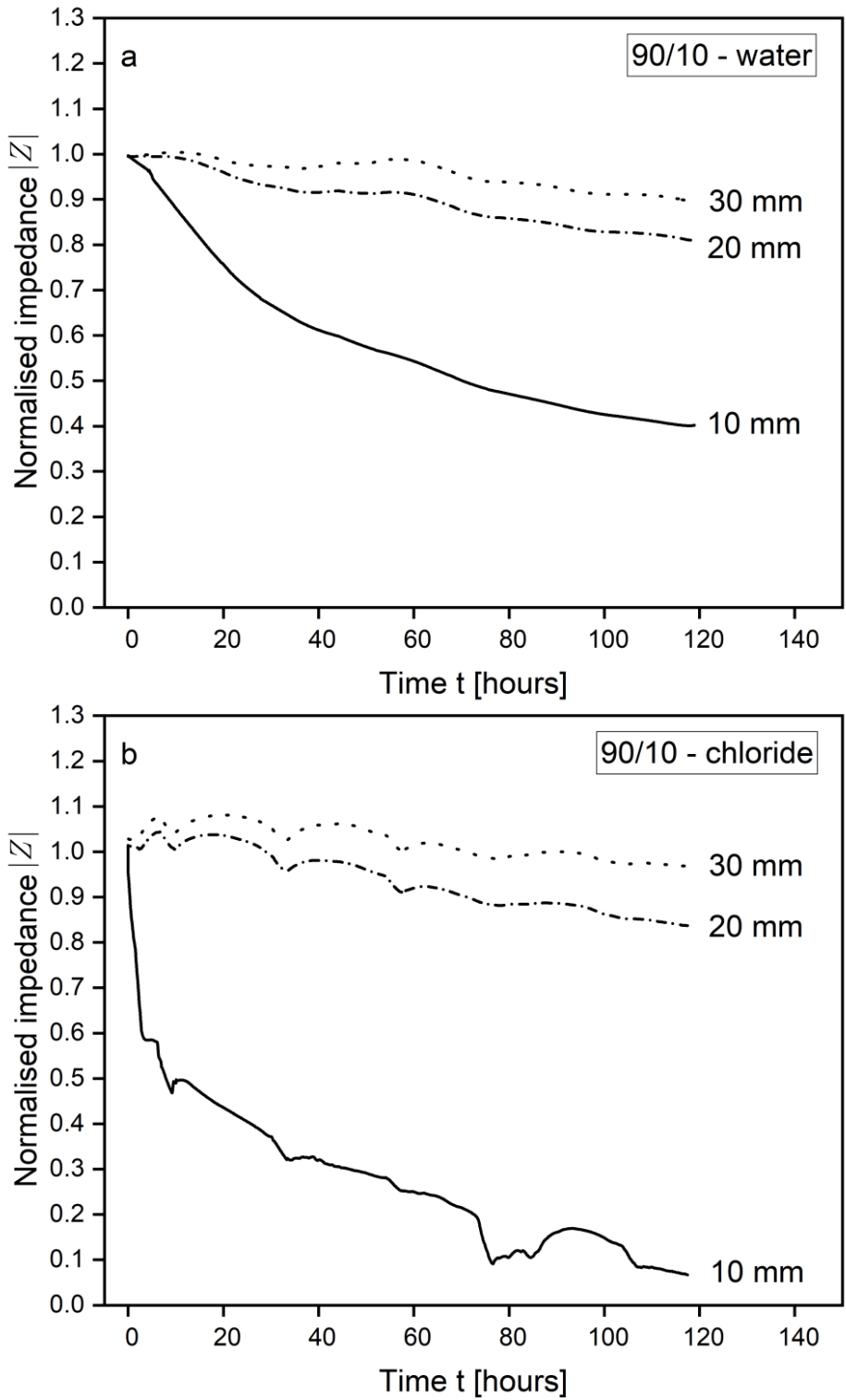

*Figure 6: Normalized single-frequency two-point impedance between the stainless steel rods embedded in 90/10 wt. % MgO/HY samples cured at 98% RH as a function of the exposure time and cover depth, relative to the sample base. Figure 6a displays the measurements of the samples exposed to tap water and Figure 6b displays the measured impedances for the sample immersed in 0.2 M KOH + 3 % NaCl.*

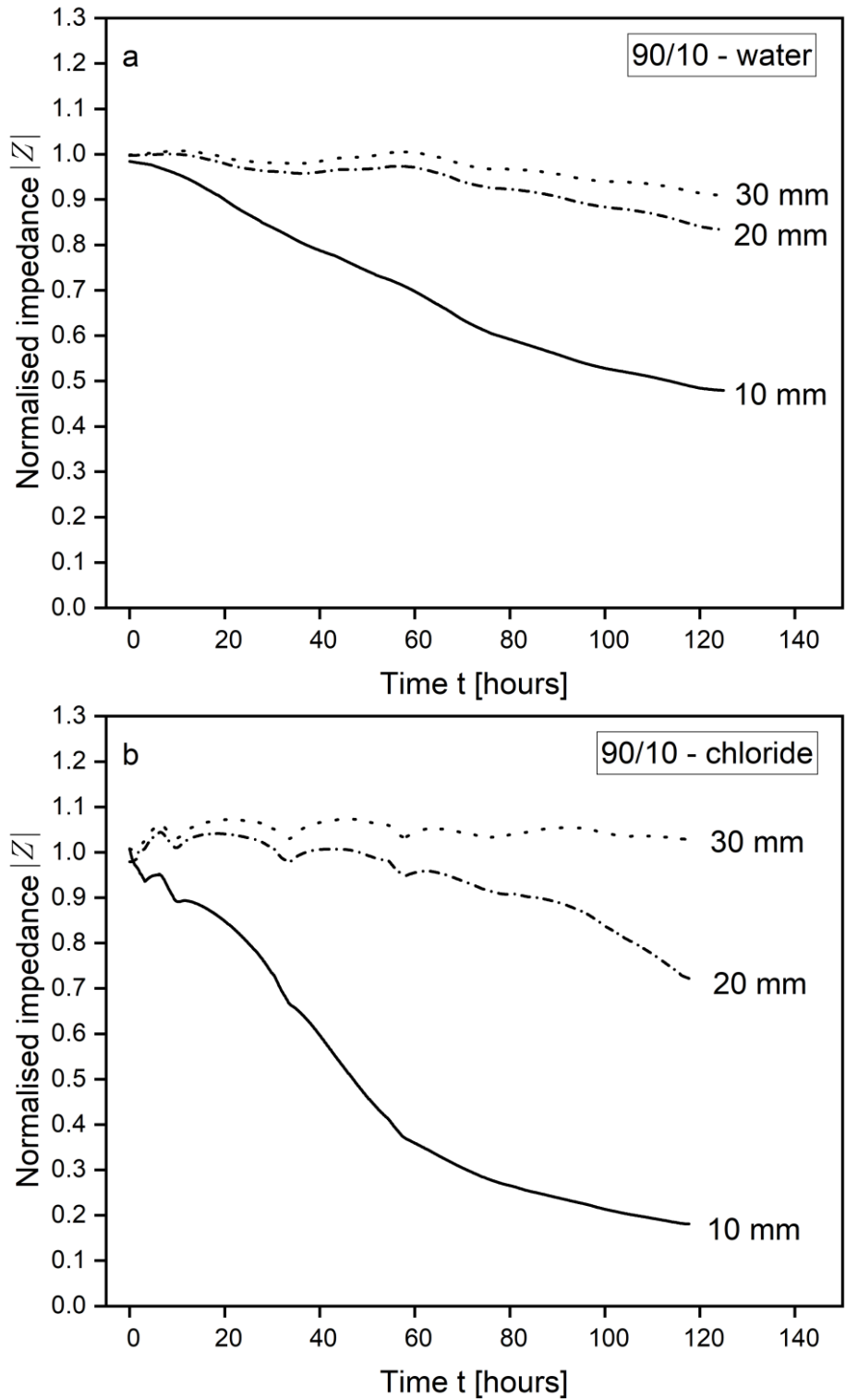

*Figure 7: Normalized single-frequency two-point impedance between the stainless steel rods embedded in 90/10 wt. % MgO/HY samples cured at 57% RH as a function of the exposure time and cover depth, relative to the sample base. Figure 7a displays the measurements of the samples exposed to tap water and Figure 7b displays the measured impedances for the sample immersed in 0.2 M KOH + 3 % NaCl.*

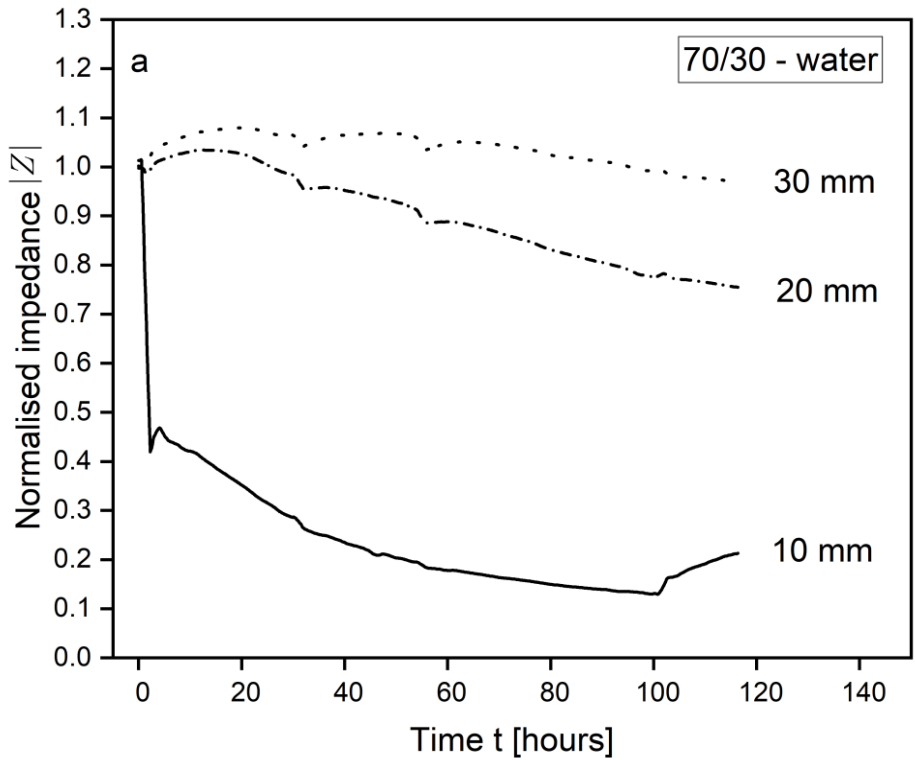

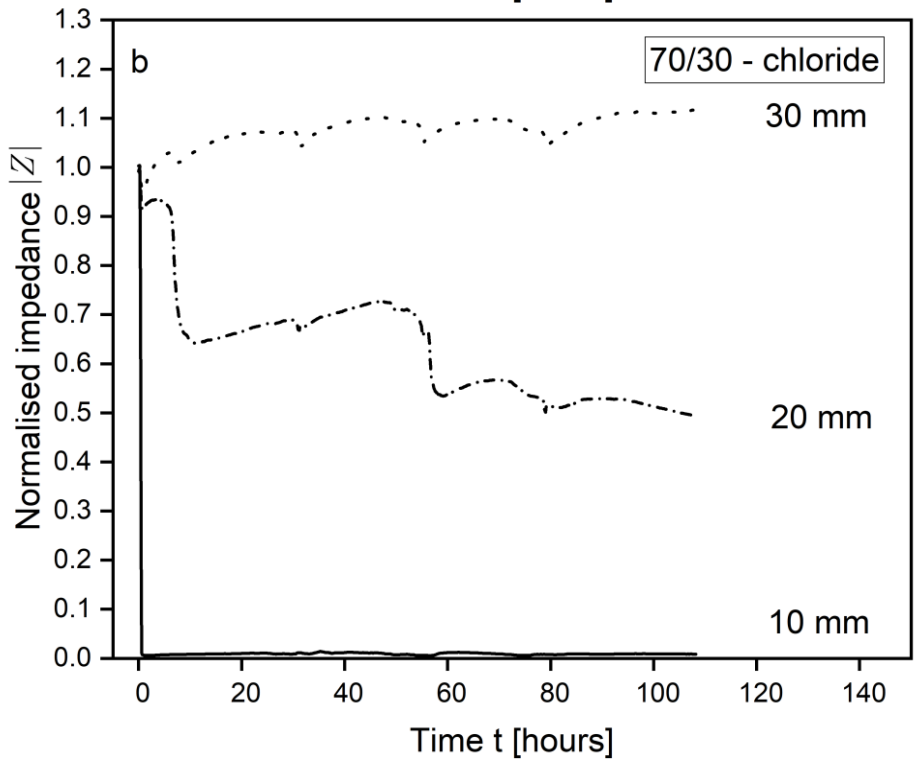

*Figure 8: Normalized single-frequency two-point impedance between the stainless steel rods embedded in 70/30 wt. % MgO/HY samples cured at 98% RH as a function of the exposure time and cover depth, relative to the sample base. Figure 8a displays the measurements of the samples exposed to tap water and Figure 8b displays the measured impedances for the sample immersed in 0.2 M KOH + 3 % NaCl.*

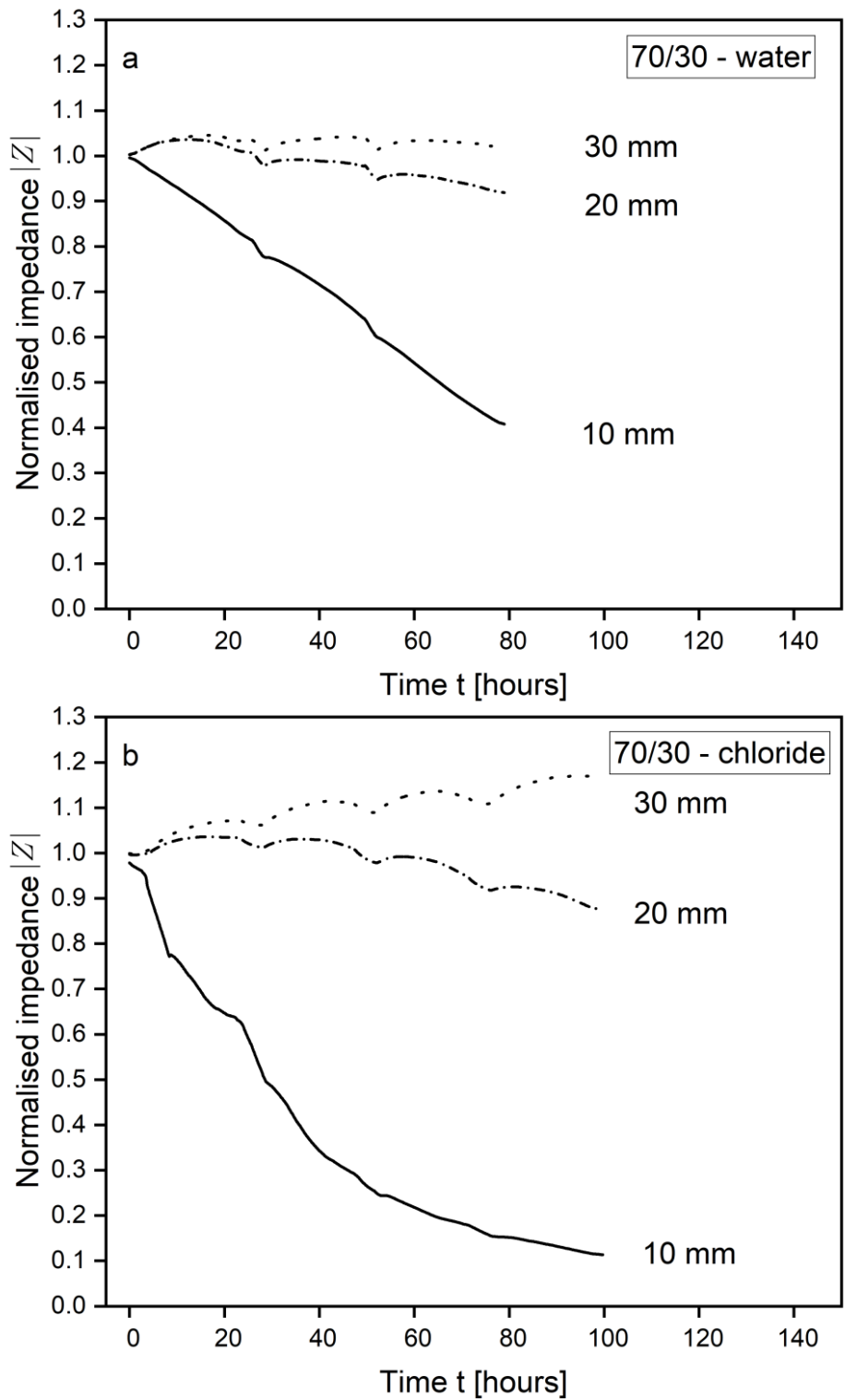

*Figure 9: Normalized single-frequency two-point impedance between the stainless steel rods embedded in 70/30 wt. % MgO/HY samples cured at 57% RH as a function of the exposure time and cover depth, relative to the sample base. Figure 9a displays the measurements of the samples exposed to tap water and Figure 9b displays the measured impedances for the sample immersed in 0.2 M KOH + 3 % NaCl.*

The observed decrease was close to instantaneous for samples cured at 98 % RH (Figures 8a and 8b), reaching values of 40 % and 2 %, respectively, relative to the initial single-frequency impedance within 2 hours. Samples cured at 57% RH (Figures 9a and 9b) approached the same relative impedance, but the recorded reduction was significantly more gradual. Analogous to the 90/10 samples, steel rods embedded at a cover depth of 30 mm appeared unaffected by the capillary uptake of both tap water and concentrated chloride solution. At the intermediate cover depth of 20 mm, the reduction was more pronounced for the 70/30 samples than for the 90/10 ones in similar exposure and curing conditions.

As shown in Figures 10 to 13, the corrosion current densities of carbon steel embedded MgO/HY blends were generally low and within the orders of $1 \times 10^{-9}$ to $1 \times 10^{-7}$ A/cm². Irrespective of the curing and exposure conditions, the instantaneous corrosion current density of the carbon steel rods embedded in the 90/10 mortar blend plateaued at $(2.0 + 0.5) \times 10^{-8}$ A/cm² after 2 to 3 days at a cover depth of both 20 and 30 mm. Analogous to the ingress of moisture, the initial increase in $i_{corr}$ was more pronounced for the rebars closest to the water level. Prisms cured at 98% RH reached a plateauing current density of $(2.5 \pm 0.5) \times 10^{-7}$ A/cm² for both the tap water and alkaline, chloride containing test solution (Figures 10a and 10b). Samples cured at 57% RH featured a slightly lower corrosion rate of $(9.0 \pm 1.0) \times 10^{-8}$ A/cm² under the ingress of water at 10 mm (Figure 11a). Upon removing the sample from the test solution, the corrosion rates decreased by a factor of 2 for all rebars exposed to the ingress of water as well as the carbon steel rods exposed to chloride-containing solution at the lowest cover depth. As elucidated in the description of the two-point impedance results, mortar blends made from 70/30 MgO/HY appeared to take up chloride-containing solution more readily than the 90/10 blends. Correspondingly, the measured corrosion current densities were the highest for both of the 70/30 samples immersed in 0.2 M KOH + 3 % NaCl. As illustrated in Figure 12b and Figure 13b, the maximum instantaneous corrosion rate measured for the 98 % and 57 % RH sample amounted to $4 \times 10^{-7}$ A/cm² and $5 \times 10^{-7}$ A/cm² after 3 days, respectively. Even though the single-frequency impedance measurements suggest an equally rapid uptake of water into the 70/30 samples (Figure 8a and 9a), the measured corrosion rates (Figure 12a and 13a) fell short of those recorded in the 90/10 sample blends. Apart from the observed order of magnitude increase in various $i_{corr}$ at the lowest cover depth (10 mm) compared to the remaining cover depths (20 and 30 mm) upon exposure to water, no clear trend can be established with respect to the sample composition, test solution or curing condition. It must be noted that the observed corrosion rates approach the lower limit at which the LPR method can reliably quantify instantaneous corrosion rate (0.1 µA/cm² or less) [28]. For this reason, we do not consider the differences in the measured $i_{corr}$ across various sample compositions and curing conditions at such low levels significant.

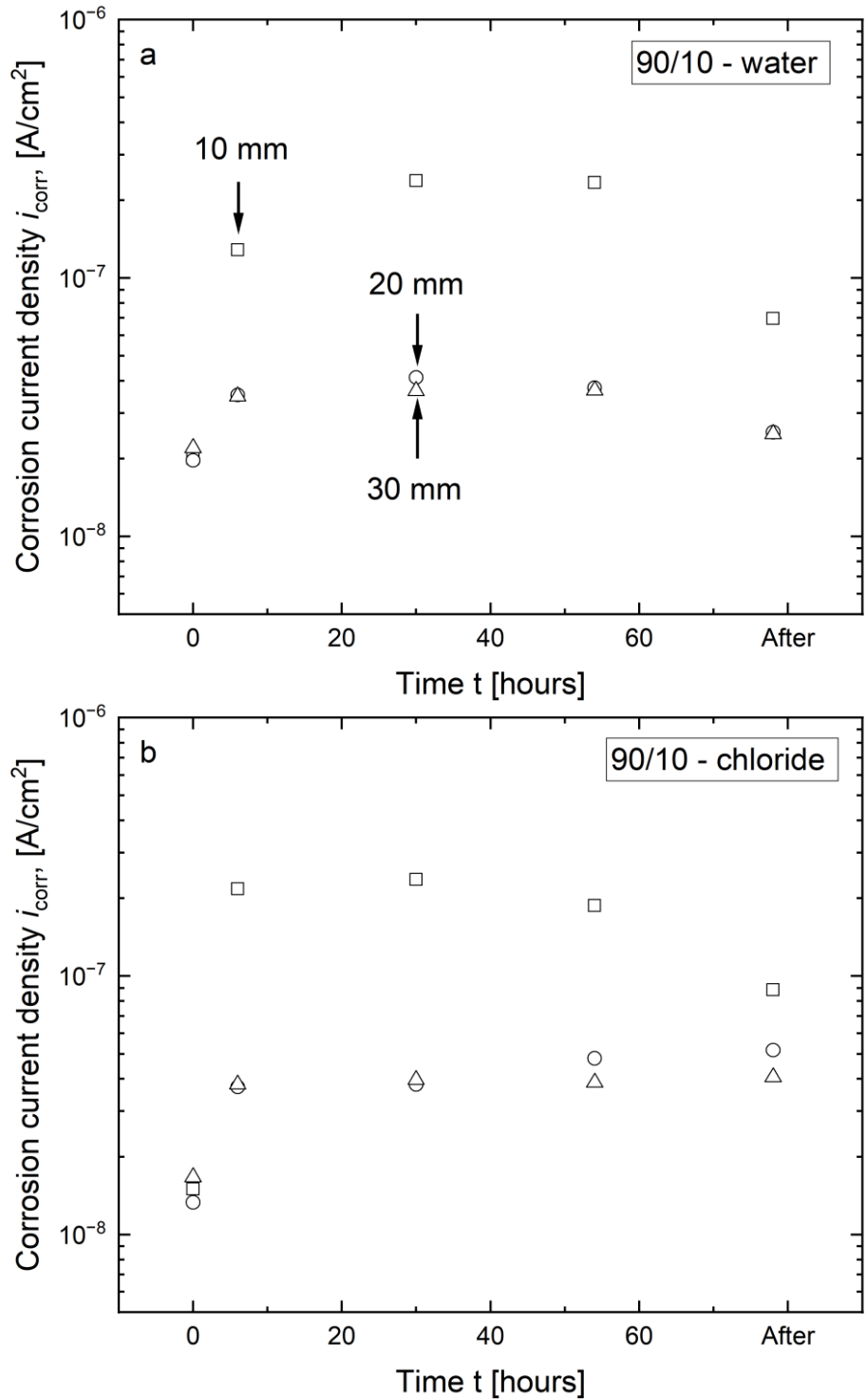

*Figure 10: Corrosion current densities of carbon steel rods embedded in 90/10 wt. % MgO/HY samples cured at 98% RH as a function of the exposure time and cover depth, relative to the sample base. Figure 10a displays the measurements of the samples exposed to tap water and Figure 10b displays the measured impedances for the sample immersed in 0.2 M KOH + 3 % NaCl.*

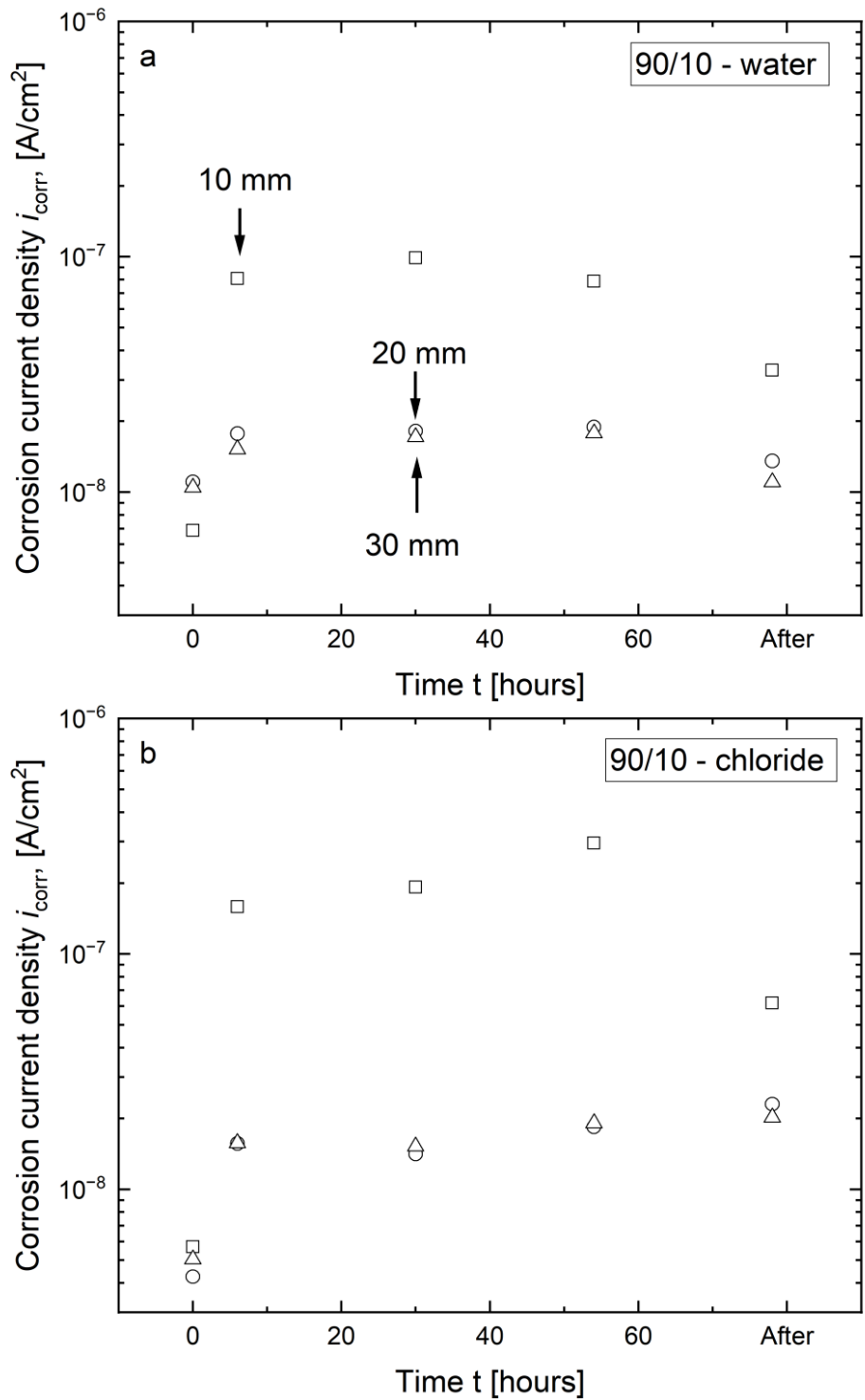

*Figure 11: Corrosion current densities of carbon steel rods embedded in 90/10 wt. % MgO/HY samples cured at 57% RH as a function of the exposure time and cover depth, relative to the sample base. Figure 11a displays the measurements of the samples exposed to tap water and Figure 11b displays the measured impedances for the sample immersed in 0.2 M KOH + 3 % NaCl.*

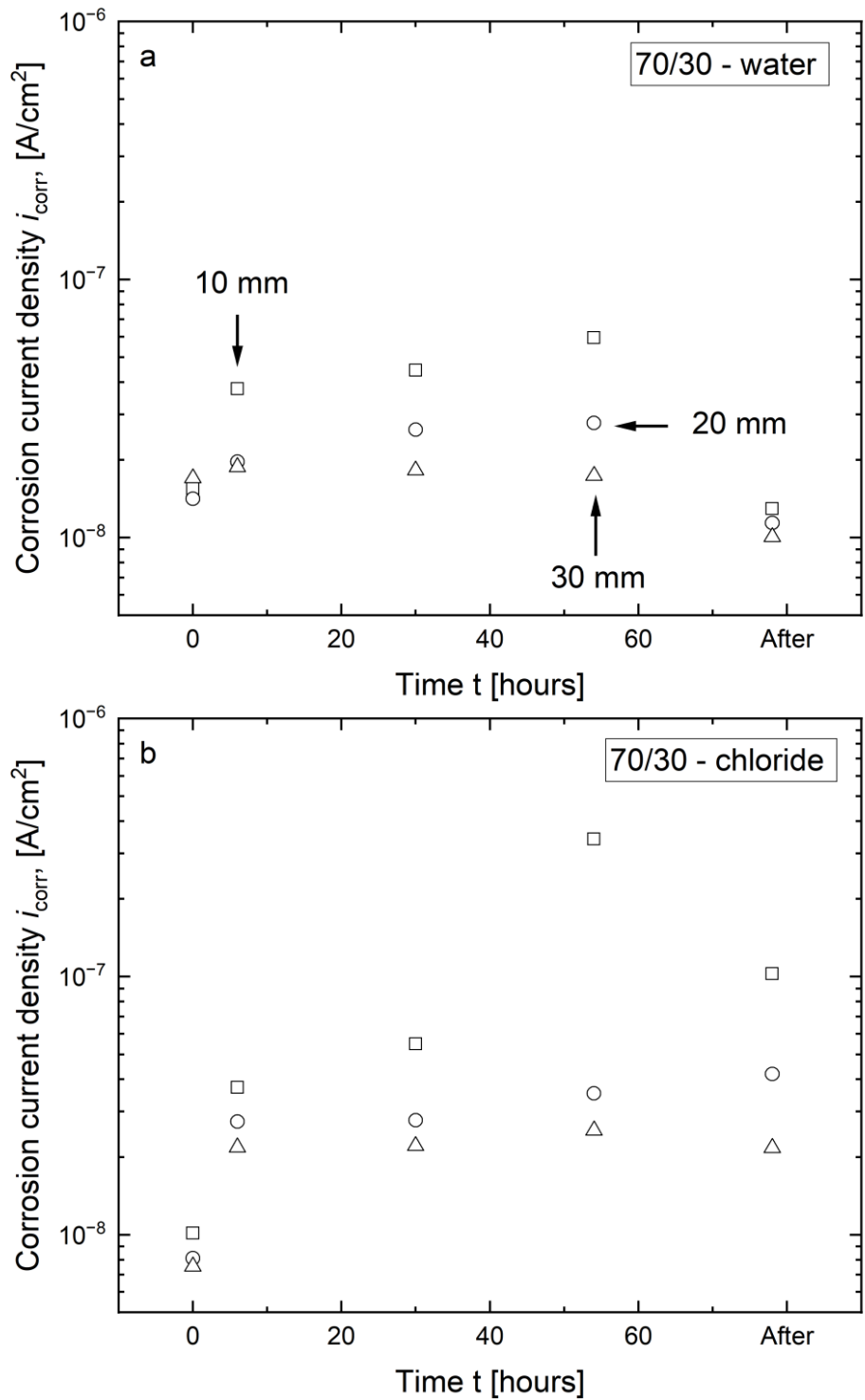

*Figure 12: Corrosion current densities of carbon steel rods embedded in 70/30 wt. % MgO/HY samples cured at 98% RH as a function of the exposure time and cover depth, relative to the sample base. Figure 12a displays the measurements of the samples exposed to tap water and Figure 12b displays the measured impedances for the sample immersed in 0.2 M KOH + 3 % NaCl.*

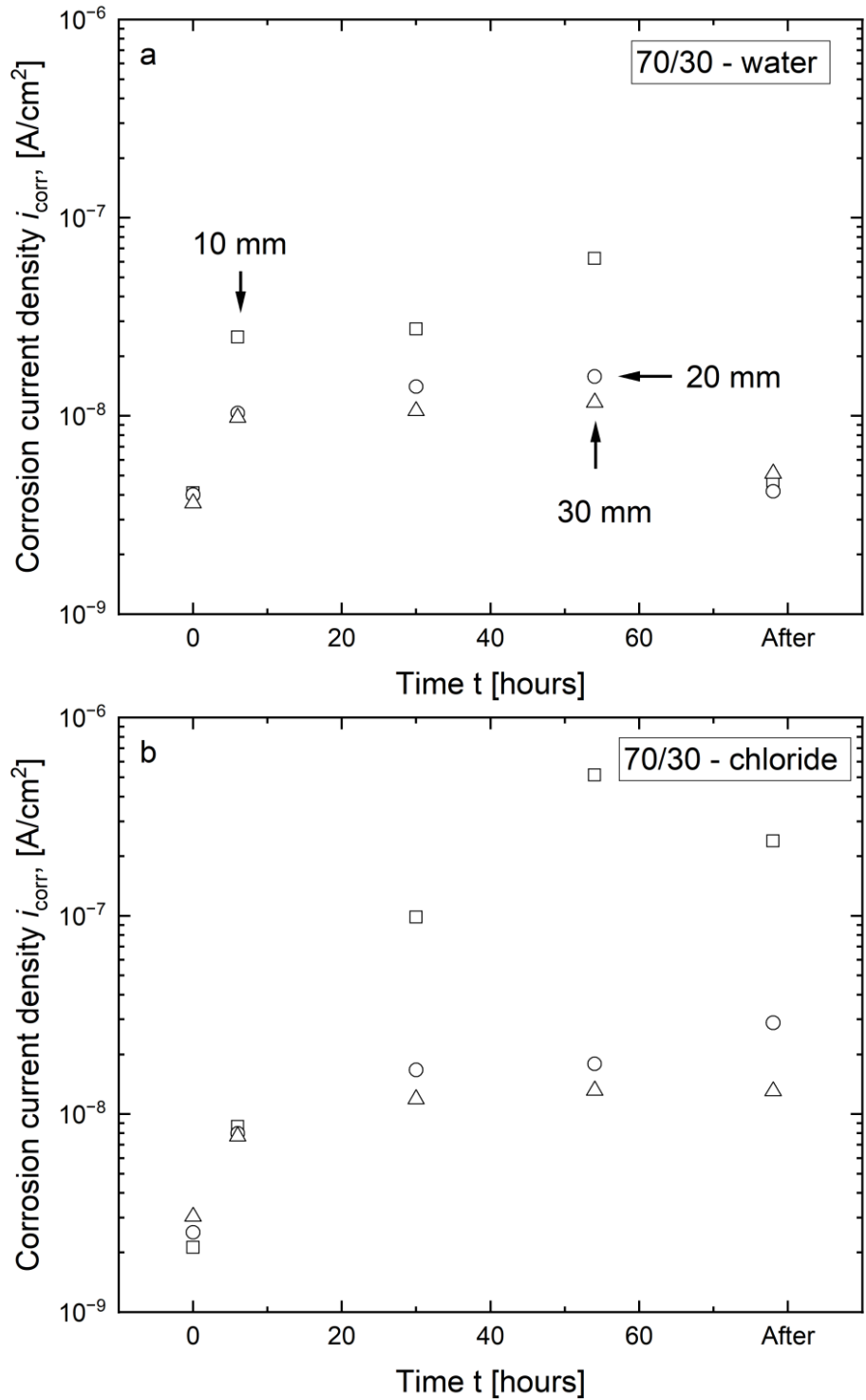

*Figure 13: Corrosion current densities of carbon steel rods embedded in 70/30 wt. % MgO/HY samples cured at 57% RH as a function of the exposure time and cover depth, relative to the sample base. Figure 13a displays the measurements of the samples exposed to tap water and Figure 13b displays the measured impedances for the sample immersed in 0.2 M KOH + 3 % NaCl.*

# 4 Discussion

## 4.1 Phase analysis of pastes

Formation of Mg-chlorides was not observed for 90/10 or 70/30 pastes cured in an alkaline chloride solution for 28 d by XRD or TGA. Possibly, chloride concentration, curing time or penetration depth of the alkaline chloride solution were not sufficient to form Mg-chlorides at the chosen experimental conditions. The presence of chlorides in the curing solution led to a shift of the HCB decomposition peak between 300 and 430°C towards lower temperatures, whereas in alkaline, chloride-free solution, the thermal decomposition behavior of HCB remained unaffected. A shift of the decomposition peak due to change in crystallite size can be excluded, as a recrystallization process would involve the formation of larger crystallites, leading to a shift towards higher rather than lower temperatures. The shift of the HCB peak could therefore indicate a limited uptake of chlorides by this phase.

Besides the shift of the HCB peak in TGA, a decrease of gel water was observed for 90/10 and 70/30 pastes cured in alkaline solution (free of chlorides) and in the alkaline chloride solution. The decrease of gel water was not linked to the shift of HCB's decomposition peak, since the shift of the peak was observed only for pastes cured in alkaline chloride solution. A possible explanation for the loss of gel water could be the high pH of both curing solutions, which could have affected the lime/carbonic acid equilibrium and through this mechanism also the carbonate incorporated into HCB during precuring. The decrease of gel water could therefore indicate a transformation of some of the HCB to brucite, resulting in the release of carbonate.

In MgO/HY concrete, further phase changes may occur, including the formation of soluble or expansive Mg-chlorides such as bischofite ($MgCl_2 \cdot 6H_2O$) or phases typical for Mg-oxychloride cements (Sorel phases) at higher chloride concentrations [14]. Whilst Mg-oxychloride cements are generally deemed incompatible with steel reinforcements due to their low pore solution pH and high chloride concentrations [15], the formation of bishofite or Sorel phases in MgO/HY concrete may be positive with regard to the initiation of chloride-induced reinforcement corrosion. Under the continual exposure to chloride-rich environments, these phase changes may actively buffer the free chloride content in the concrete and prevent the accumulation of chlorides at the steel-concrete-interface. As the MgO/HY binder additionally features a higher pore solution pH, the prevailing $Cl^-/OH^-$ ratio at the steel concrete interface would be lower, thereby reducing the probability of chloride-induced reinforcement corrosion [29]. No results about the durability of MgO/HY binder exposed to chloride-rich environments have thus far been reported in the scientific literature.

## 4.2 Rapid chloride ingress test

The rapid chloride ingress test performed on 90/10 and 70/30 mortars after SIA 262/1 Appendix B showed a high penetration resistance against chlorides. Both mortar mixes showed comparable chloride penetration resistances, indicating low permeability of the mortars despite different binder compositions. However, the small penetration depths obtained by the rapid chloride ingress tests do not necessarily provide information on the potential of corrosion of rebars in steel-reinforced MgO/HY concrete. The standard SIA 262/1 Appendix B has been initially developed for concrete, which has a lower paste volume than mortars. Furthermore, the standard has been designed to link chloride penetration depths to risk of corrosion in Portland cement-based concrete. Due to the change of the binder, it cannot be guaranteed that the relationship between chloride penetration and corrosion will remain valid for the MgO/HY binder. The rapid chloride ingress test rather confirms a low permeability for the analyzed mortar mixes, which prevents fast chloride penetration under the specific testing conditions.

The SIA 262/1 Appendix B standard further employs highly alkaline test solutions (0.2 M KOH). Whilst the pH of both test solutions, one additionally containing 3 % NaCl and the other one being chloride-free, is intended to mimic the pore solution pH of Portland cement-based binders, it is not representative of the MgO/HY binder investigated in this study. With regard to future investigations into the chloride permeability of Mg-based and other low pH binders, it may thus be worthwhile considering modifying the standardized rapid chloride ingress test and adjusting the test solution pH to the prevailing degree of alkalinity in the binder investigated. As the accelerated ingress test conducted in this study is further not representative of the ingress of chlorides under natural conditions, it is recommended that the here presented preliminary investigation is complemented by other tests that emulate the ingress of chlorides due to bulk diffusion and advection, e.g. in cyclic wet/dry exposure.

## 4.3 Corrosion experiments

The electrochemical measurements presented in this study further demonstrate a good short-term corrosion performance of both mortar compositions studied. Single-frequency impedance measurements show that the capillary ingress of water and chloride-containing test solution occurs more readily in the 70/30 mortar samples compared those made from 90/10 MgO/HY. These differences likely relate to differences in the porosity of both mortar blends. As evident from a significant reduction in the normalized impedance measurements over time, the pore network in the vicinity of the steel rods embedded at a cover depth of 10 mm has become increasingly more saturated over the course of the experiment. Despite the continued ingress of moisture, their measured instantaneous corrosion rates do not exceed $3.5 \times 10^{-7}$ A/cm$^2$ (compare Figure 9b, 54 hours), equivalent to a cross-sectional loss of less than 1 μm/year. As elucidated in the previous section, various other corrosion rates measured at the rebars embedded at a cover depth of 20 and 30 mm flatten out after 2 to 3 days, reaching a negligible steady-state cross-sectional loss of less than 1 μm/year. At either the intermediate or the highest cover depth, there is no significant difference between the measured $i_{corr}$ in samples exposed to tap water and those in contact with concentrated chloride solution. The linear polarization resistance measurements are thus in line with the results of the rapid chloride ingress test (see Section 4.2). The apparent similarity between the corrosion current densities measured in the presence and absence of chlorides raises the question as to whether the exposure time was sufficiently long to allow for chloride-induced corrosion to initiate. With regard to future investigations into the corrosion performance of the MgO/HY binder, it is recommended to repeat the here presented ingress tests using a test solution with a lower pH, i.e. at lower Cl$^-$/OH$^-$ ratio, to facilitate corrosion initiation [29].

The pore solution pH of MgO-based binders is typically buffered at pH values ranging from 10.5 to 11.0 [3, 30]. In comparison, the pore solution of uncarbonated Portland cement is buffered well above a pH of 12.5 [31]. This high degree of alkalinity, and correspondingly, the thermodynamically favorable stabilization of iron (hydr)oxide phases on the reinforcement surface, is the primary reason as to why the corrosion rate of steel in concrete is negligible [32]. The apparent discrepancy in the degree of alkalinity between these novel and traditional cementitious binders has recently given rise to the presumption that MgO-based binders may be unsuited for reinforcement applications [30]. Despite these theoretical considerations, the measured corrosion rates of carbon steel embedded in MgO/HY mortars presented in this study are comparable to the corrosion rate of passive steel embedded in Portland cement [33]. Figure 14 displays the Pourbaix diagram of iron, together with the open circuit potential (OCP) measurements obtained during the corrosion rate measurements of carbon steel embedded in the MgO/HY binder, exposed to the ingress of water.

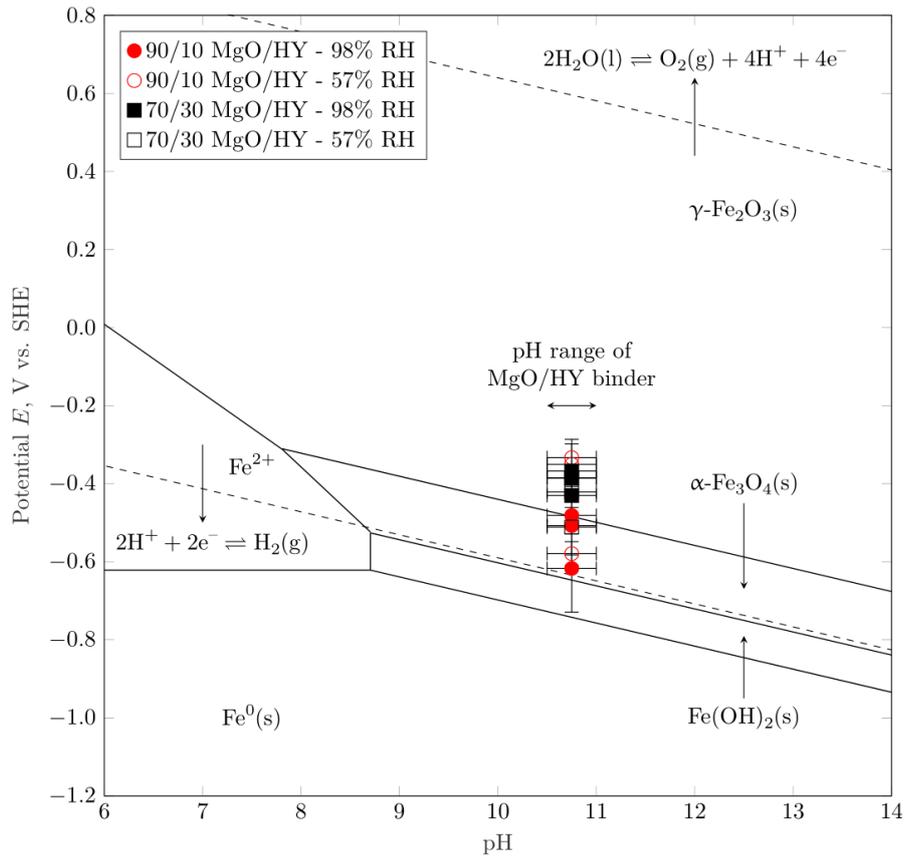

*Figure 14: Pourbaix diagram of iron, including selected solid iron (hydr)oxide phases, together with the open circuit potential (OCP) measurements obtained during the corrosion rate measurements presented in this study. The diagram was created using the open source Python library pourPy [34] and the thermodynamic data published in Furcas et al. [35]. The error bar in x (pH) and y (potential) direction represent the pH range of MgO/HY binders and the standard deviation of 3 OCP measurements for each binder composition and curing condition studied.*

Across the pH range characteristic to the MgO/HY binder, various OCP measurements are well within the potential-pH stability regime of the solid iron phases white rust (Fe(OH)$_2$(s)), magnetite (α-Fe$_3$O$_4$(s)) and maghemite (γ-Fe$_2$O$_3$(s)). From a thermodynamic point of view, the corrosion of carbon steel in the investigated MgO/HY binder can thus be considered analogous to the corrosion of carbon steel in partially carbonated Portland cement-based concrete. Firstly, both systems feature a pH significantly lower than the pH of uncarbonated Portland cement-based concrete. The pore solution pH of the Mg-based binder investigated in this study is somewhere in between that of fully carbonated and uncarbonated Portland cement-based concrete. Secondly, measured potentials of carbon steel embedded in the MgO/HY binder, as well as in moist, wet or dry carbonated concrete [36], are in the passivity regime of the classical Pourbaix diagram of iron [16, 32]. Although this overlap suggests that carbon steel embedded in either cementitious system experiences negligible corrosion, it is well known that the instantaneous corrosion rate of steel in carbonated concrete can in fact be technically relevant [37]. As previously detailed in Angst et al. [22], the major influencing factor controlling the degree of corrosion-related damage in carbonated concrete is the moisture state at the steel reinforcement, and therefore, implicitly the concrete cover depth, its pore structure and its integrity (i.e., limited cracking). Similarly, the primary predictor of the corrosion rate of carbon steel embedded in the MgO/HY binder investigated in this study is found to be the mortar cover depth (compare Figures 10 to 13). The satisfactory resistance against the ingress of moisture (compare Figures 6 to 9) suggests that the mortar features a low porosity, sufficient to control the moisture transport through the mortar cover during exposure to liquid water over a duration of 2-3 days. Due to the overall negligible corrosion rates (in the order of $1 \times 10^{-7}$ to $1 \times 10^{-8}$ A/cm$^2$ at a low cover depth of 10 mm and $1 \times 10^{-8}$ to $1 \times 10^{-9}$ A/cm$^2$ at intermediary cover depths of 20-30 mm), we conclude that the corrosion rate of carbon steel in the MgO/HY binder also primarily depends on the moisture state of the steel reinforcement and that the comparatively low pore solution pH may be tolerable, as long as the steel-concrete-interface is sufficiently dry [17, 19, 22]. As highlighted previously, further experimental work is needed to assess the long-term corrosion performance of the binder as well as secondary, corrosion-related degradation processes, including cracking and spalling of the concrete cover.

# 5  Conclusion

The formation of Mg-chlorides has not been observed by for MgO/HY pastes cured in alkaline chloride or pure chloride solution for 7 and 28 d. Rapid chloride ingress tests showed high penetration resistance against chlorides of 90/10 and 70/30 MgO/HY mortars prepared with w/c of 0.50 and superplasticizer addition. Penetration depths were marginal for both mortar mixes ranging between 1-5 mm at a total sample length of 50 mm. The associated chloride migration coefficients were $D_{Cl}$ = 0.6 ± 0.3 and 1.0 ± 0.6 · 10$^{-12}$ m$^2$/s for the 90/10 and 70/30 mortars, respectively. In comparison, the maximum recommended limit of the chloride migration coefficients is 10 · 10$^{-12}$ m$^2$/s for concrete exposure classes XD2b(CH) and XD3(CH) [25]. Although these first screening results are promising, further testing is needed to establish whether the accelerated testing results hold true for natural exposure conditions including the ingress of chlorides due to bulk diffusion and advection.

LPR measurements further show that carbon steel rods embedded in 90/10 and 70/30 MgO/HY mortars exposed to the capillary ingress of water and chloride-containing solution experience negligible corrosion rates in the orders of $1 \times 10^{-9}$ to $1 \times 10^{-7}$ A/cm$^2$. Across all experiments conducted, the corrosion rate of carbon steel rods with the lowest cover depth of 10 mm, i.e. the ones close to the water table, is one order of magnitude higher than that of the rods embedded at 20 and 30 mm. Findings of these accelerated, short-term experiments suggest that, analogous to the corrosion of steel in carbonated concrete, the moisture content is the main predictor of the corrosion rate of steel in MgO/HY binders, at least as long as the matrix surrounding the steel remains chloride-free.

The tested MgO-based binders could prospectively be suitable for use in structural concrete with steel reinforcement. Further investigations are required to assess the long-term corrosion performance of MgO-based binders as well as their resistance to other exposure conditions.

## CRedit authorship contribution statement

**Fabio Enrico Furcas:** Conceptualization, Methodology, Validation, Formal analysis, Investigation, Data curation, Writing – original draft, Visualization. **Alexander German**: Conceptualization, Methodology, Validation, Formal analysis, Investigation, Data curation, Writing – original draft, Visualization. **Frank Winnefeld**: Conceptualization, Methodology, Writing – review and editing, Supervision, Funding acquisition. **Pietro Lura**: Writing – review and editing, Supervision, Funding acquisition. **Ueli Angst**: Writing – review and editing, Supervision, Funding acquisition.

## Declaration of competing interest

The authors declare no conflicts of interest.

## Acknowledgments

The authors thank the European Research Council (ERC) for the financial support provided under the European Union's Horizon 2020 research and innovation program (grant agreement no. 848794). The authors also thank Dr. Yurena Seguí Femenias (DuraMon) for the provision of the impedance data loggers. Boris Ingold (Empa) is acknowledged for performing the rapid chloride migration test.